# Time delay of slow electrons by a diatomic molecule described by non-overlapping atomic potentials model

M. Ya. Amusia[1,2] and A. S. Baltenkov[3]

[1] *Racah Institute of Physics, the Hebrew University, Jerusalem, 91904 Israel*
[2] *Ioffe Physical-Technical Institute, St. Petersburg, 194021 Russia*
[3] *Arifov Institute of Ion-Plasma and Laser Technologies*,
*Tashkent, 100125 Uzbekistan*

**Abstract.** We study the elastic scattering of slow electrons by two-atomic molecule in the frame of non-overlapping atomic potentials model. The molecular continuum wave function is represented as a combination of a plane wave and two spherical *s*-waves, generated by the centers of atomic spheres. The asymptotic of this function determines in closed form the amplitude of elastic electron scattering. We show that this amplitude cannot be represented as a series of spherical functions. Therefore, it is impossible to use straightly the usual S-matrix methods to determine the scattering phases for non-spherical targets. We show that far from molecule the continuum wave function can be presented as an expansion in other than spherical orthonormal functions. The coefficients of this expansion determine the molecular scattering phases for non-spherical molecular systems. In such an approach, we calculate the Wigner times delay for slow electron scattered by two-atomic target.

## 1. Introduction

Experiments with atomic photoionization under the action of attosecond laser pulses provided qualitatively new information on the real time-dynamics of this process (see [1-10] and references therein). The accepted interpretation of these experimental data relies on the time delays originally introduced by Eisenbud, Wigner, and Smith [11-13]. The time delay, as a quantum dynamical observable, is extremely sensitive to the phase shifts of photoelectron wave function, or, more precisely, to its energy derivative. As compared to atomic photoionization, the molecular time-delay presents a more difficult for description phenomenon due to the anisotropic nature of the molecular scattering potential. The theoretical approaches to time-dynamic photoionization of the simplest molecular system (hydrogen ion $H_2^+$), for which in spheroidal coordinates the separation of variables in the wave equation is possible [14], were developed in papers [15-17]. Temporal analysis of more complex molecules photoionization leads to so-called *angular time delay* [18-23] as a function depending on the orientation of photon polarization **e** and photoelectron momentum **k** vectors, relative to the molecular axes $\{\mathbf{R}_i\}$.

Paper [24], for example, presents special features of theoretical and computational techniques that are usually employed in describing the molecular continuum wave-functions and phase shifts. The complex of mathematical programs *ePolyScat* [25, 26] (see Lucchese Group Website, 2015) implements numeric realization of these ideas. Briefly, the quote on page 361 of review [24] describes the main elements of this method: "…all the quantities, such as the bound and the continuum orbitals, interaction potentials, *etc*., are expanded around a single center, normally the center-of-mass (COM) of the system. Then, the scattering equations for the continuum electron function will be *similar to those for electron-atom scattering* and will become computationally more tractable". Let us pay attention to the final part of this quotation, according to which the developed calculation method is mainly due to the convenience of calculations.

In the description of another so-called continuum multiple scattering CMS-method [27] we find: "In the electron scattering case, *the continuum electron is treated as moving in a spherically averaged molecular field* which is divided into an inner region, chosen from the electron densities of the isolated component atoms, and an outer region encompassing the whole molecule". Thus, outside the molecular sphere, the field of the molecule is considered to be spherical, and the molecular phases of scattering are determined by the conditions of matching the wave function of the continuum on the surfaces of atomic and molecular spheres[1].

These methods assume that the asymptotic form of the wave function far from the molecule (similar to atomic case) is a sum of a plane wave plus a single spherical wave (SSW) emitted by the molecular center

$$\psi_{\mathbf{k}}^{+}(\mathbf{r} \to \infty) = e^{i\mathbf{k}\cdot\mathbf{r}} + F(\mathbf{k},\mathbf{k}')\frac{e^{ikr}}{r}. \qquad (1)$$

Here **r** is the radius-vector of scattering electron in the coordinate system originating at the target's center of mass, **k** and $\mathbf{k}' = k\mathbf{r}/r$ are the initial and final directions of the continuum electron and $F(\mathbf{k},\mathbf{k}')$ is the elastic scattering amplitude

$$F(\mathbf{k},\mathbf{k}') = \frac{2\pi}{ik}\sum_{l,m}(e^{2i\eta_l} - 1)Y_{lm}^{*}(\mathbf{k})Y_{lm}(\mathbf{k}'). \qquad (2)$$

Here $\eta_l$ are the molecular phase shifts. Spherical functions in (2) are $Y_{lm}(\mathbf{r}) \equiv Y_{lm}(\vartheta,\varphi)$, where $\theta$ and $\varphi$ are the polar and azimuthal angles of **r** and similar for $Y_{lm}(\mathbf{k}) \equiv Y_{lm}(\vartheta_k,\varphi_k)$. Hence, in [24-27] the solution of the problem of electron scattering by a non-spherically symmetric potential is reduced, without sufficient grounds, to the usual S-matrix method of the partial waves for a spherical target. These recipes to build a molecular continuum wave function (and molecular phase shifts [20-23]) are considered as a matter-of-course and, as far as we know, are beyond any doubt.

In spite of the general believe let us analyze the correctness of this approach to the calculation of the molecular continuum wave functions and molecular scattering phases. It is reasonable to consider for this aim a model molecular system that allows an exact solution/ Then using this example, to verify the correctness of the above mentioned approaches for calculating molecular phases and Wigner electron delay times by polyatomic systems. In the present paper, we shall do so on example of the target formed by two spatially separated by distance **R** non-overlapping short-range atomic potentials.

The problem of elastic scattering of a particle on a non-spherical target and S-matrix method for this case was developed in [29], where it was shown that, asymptotically at great distances from the molecule the continuum wave function one can present as an expansion in a set of other, than spherical $Y_{lm}(\mathbf{k})$, orthonormal functions $Z_\lambda(\mathbf{k})$ depending, in particular, on the number of atoms forming the target and on the relative orientation of the scattering centers in space, *etc*. In Section 2 we briefly discuss the main ideas of [29]. In section 3 we derive the main formulas for the particle elastic scattering by two-center molecular system. The wave function of the molecular continuum is represented as a combination of a plane wave and two spherical *s*-waves generated by each of the centers. The amplitude of elastic scattering for this system $F(\mathbf{k},\mathbf{k}',\mathbf{R})$ is represented in a closed form, but not in the form of an S-matrix expansion in the partial waves. The expansion of function $F(\mathbf{k},\mathbf{k'},\mathbf{R})$ in the series of $Y_{lm}(\mathbf{k})$- and $Z_\lambda(\mathbf{k})$-functions is performed with the aim to obtain general expressions for molecular phase shifts $\eta_\lambda(k,R)$. In Section 4 we use the obtained molecular phases to calculate the Wigner times delay in the process of slow electron scattering by two-atomic molecule. Section 5 is the Conclusion.

---

[1] We have demonstrated quite a while ago [28] that the introduction of a molecular sphere into the muffin-tin-potential model leads to some non-physical features in the molecular continuum wave functions.

## 2. Method of partial waves for non-spherical targets [29]

It is known that the wave function describing elastic scattering of a particle by a spherically symmetrical potential is defined by the expression [30]

$$\psi_\mathbf{k}^+(\mathbf{r}) = 4\pi \sum_{l=0}^{\infty} R_{klm}(r) Y_{lm}(\mathbf{r}) Y_{lm}^*(\mathbf{k}), \tag{3}$$

where the radial part of the wave function has the asymptotic form

$$R_{klm}(r \to \infty) \approx e^{i(\delta_l + \frac{\pi l}{2})} \frac{1}{kr} \sin(kr - \frac{\pi l}{2} + \delta_l). \tag{4}$$

A molecular potential as a cluster of non-overlapping spherical potentials centered at the atomic sites is non-spherical. The solution $\psi_\mathbf{k}^+(\mathbf{r})$ of the Schrödinger equation with this potential is impossible to present at an arbitrary point of space in the form (3), i.e. as an expansion in spherical functions $Y_{lm}(\mathbf{r})$. However, asymptotically at great distances from the molecule the wave function can be presented as an expansion in a set of other orthonormal functions $Z_\lambda(\mathbf{k})$:

$$\psi_\mathbf{k}^+(\mathbf{r} \to \infty) \approx 4\pi \sum_\lambda R_{k\lambda}(r) Z_\lambda(\mathbf{r}) Z_\lambda^*(\mathbf{k}) \tag{5}$$

with the radial part of the wave function determined by the following expression

$$R_{k\lambda}(r \to \infty) \approx e^{i(\eta_\lambda + \frac{\pi}{2}\omega_\lambda)} \frac{1}{kr} \sin(kr - \frac{\pi}{2}\omega_\lambda + \eta_\lambda). \tag{6}$$

Here the index $\lambda$ numerates different partial functions similar to the quantum numbers $l$ and $m$ for the central field; $\omega_\lambda$ is the quantum number that is equal to the orbital moment $l$ for the spherical symmetry case; $\eta_\lambda(k)$ are the "proper molecular phases"[29]. The explicit form of functions $Z_\lambda(\mathbf{k})$ depends upon the specific type of the target field, particularly on the number of atoms forming the target and on mutual disposition of the scattering centers in space, *etc*. The functions $Z_\lambda(\mathbf{k})$, like the spherical functions $Y_{lm}(\mathbf{k})$, create an orthonormal system and therefore one has

$$\int Z_\lambda^*(\mathbf{k}) Z_\mu(\mathbf{k}) d\Omega_k = \delta_{\lambda\mu}. \tag{7}$$

The scattering amplitude for a non-spherical target, according to [29], is given by the following expression

$$F(\mathbf{k},\mathbf{k'}) = \frac{2\pi}{ik} \sum_\lambda (e^{2i\eta_\lambda} - 1) Z_\lambda^*(\mathbf{k}) Z_\lambda(\mathbf{k'}). \tag{8}$$

The total elastic scattering cross section, i.e. the cross section integrated over all directions of momentum of the scattered electron **k'** relative to vector **k**, is defined by the formula

$$\sigma(\mathbf{k}) = \frac{(4\pi)^2}{k^2} \sum_{\lambda} |Z_{\lambda}(\mathbf{k})|^2 \sin^2 \eta_{\lambda}. \quad (9)$$

Of course, this cross section depends on the mutual orientation of incident electron momentum **k** and molecule axes {**R**$_i$}. The cross section averaged over all the directions of momentum of incident electron **k** is connected with the molecular phases $\eta_{\lambda}(k)$ by the following expression

$$\overline{\sigma}(k) = \frac{4\pi}{k^2} \sum_{\lambda} \sin^2 \eta_{\lambda}. \quad (10)$$

In the case of a spherical symmetric target the formula (10) coincides exactly with the known formula for the total scattering cross section. Indeed, in the case of the central field the index $\lambda$ is replaced by the quantum numbers $l$ and $m$. But the phase of scattering by a central field is independent upon the magnetic number $m$. Therefore for the given value of the orbital moment $l$ it is necessary to sum over all $m$. This results in the factor $(2l+1)$ under the summation sign in (10).

The partial wave (6) and molecular phases $\eta_{\lambda}(k)$ are classified, according to [29], by their behavior at low electron energies, i.e. for $k \to 0$. In this limit the particle wavelength is great as compared to the target size and the function $Z_{\lambda}(\mathbf{k})$ tends to some spherical function $Y_{lm}(\mathbf{k})$. The corresponding phase is characterized in this limit by the following asymptotic behavior: $\eta_{\lambda}(k \to 0) \approx k^{2\lambda+1}$.

We apply the formulas of this Section to calculate the elastic electron scattering by a target formed by two non-overlapping atomic potentials.

### 3. Target as a cluster of two non-overlapping atomic spheres

Consider the scattering of a slow electron by two identical non-overlapping atomic potentials with the centers at **r**$_{1,2}$ = ±**R**/2. This simplest multicenter system is a good solvable example with analytical solution. It can serve as a touchstone to analyze correctness of different molecular phase shifts calculation methods.

Let us at first consider the electron continuum wave function in the field of a single atomic sphere located at the origin of coordinates. *Beyond the atomic sphere* the radial part of electron wave function is a linear combination of the usual spherical Bessel functions [31]. The coefficients of this linear combination are defined by the *atomic phase shift* $\delta_l(k)$. Here $l$ is the orbital momentum and $k$ is the electron momentum[2]. Suppose that all phases of electron scattering by a given atomic sphere, except the $s$-phase, are equal to zero. In this case the electon continuum wave function has form

$$\psi_{\mathbf{k}}^{+}(\mathbf{r}) \propto \sin(kr + \delta_0), \text{ for } r \geq \rho, \quad (11)$$

---

[2]Atomic units are used throughout this paper.

where $\rho$ is radius of the potential sphere. Assume that we know the phase $\delta_0(k)$ from the wave equation solution for isolated atom. Following to ideas [32, 33], let us extend the wave function in the form (11) to *all radiuses r*, i.e. inside the atomic sphere also, and find the boundary conditions that should be imposed on the wave function at the point $r = 0$ so that beyond the atomic sphere it would be described by formula (11). In the limit $r \to 0$, equation (11) leads to the following boundary condition that imposes to the wave function at the center of atomic sphere

$$\psi_{\mathbf{k}}^{+}(\mathbf{r} \to 0) = C\left[\frac{1}{r} + k \cot \delta_0\right] \qquad (12)$$

where $C$ is the some constant[3].

According to the Huygens-Fresnel principle, the wave function of a slow electron scattered by two non-identical atomic spheres one can present as a sum of a plane wave and two spherical *s*-waves emitted from the points $\mathbf{r}_{1,2} = \pm \mathbf{R}/2$

$$\psi_{\mathbf{k}}^{+}(\mathbf{r}) = e^{i\mathbf{k}\cdot\mathbf{r}} + D_1(\mathbf{k})\frac{e^{ik|\mathbf{r}+\mathbf{R}/2|}}{|\mathbf{r}+\mathbf{R}/2|} + D_2(\mathbf{k})\frac{e^{ik|\mathbf{r}-\mathbf{R}/2|}}{|\mathbf{r}-\mathbf{R}/2|}. \qquad (13)$$

Note, that in [34] a similar function was used to describe scattering of slow mesons by deuterons (see also [35]). Boundary conditions (12) at points $\mathbf{R}/2$ and $-\mathbf{R}/2$ for this function has the following form

$$\psi_{\mathbf{k}}^{+}(\mathbf{r})_{\mathbf{r}\to \mathbf{R}/2} \approx C_1\left[\frac{1}{|\mathbf{r}-\mathbf{R}/2|} + k\cot\delta\right]; \quad \psi_{\mathbf{k}}^{+}(\mathbf{r})_{\mathbf{r}\to -\mathbf{R}/2} \approx C_2\left[\frac{1}{|\mathbf{r}+\mathbf{R}/2|} + k\cot\delta\right]. \qquad (14)$$

Applying them to the function (13) at the points $\mathbf{r} = \pm \mathbf{R}/2$ and equating the divergent and next to divergent terms, we obtain the following relations:

$$e^{i\mathbf{k}\cdot\mathbf{R}/2} + D_1\frac{e^{ikR}}{R} + D_2 ik = D_2 k \cot\delta, \quad e^{-i\mathbf{k}\cdot\mathbf{R}/2} + D_2\frac{e^{ikR}}{R} + D_1 ik = D_2 k \cot\delta, \quad D_{2,1} = C_{1,2} \qquad (15)$$

Let us introduce the following parameters $a$, $b$, and $d$ (see formulas in [35]), using the following definitions

$$a = \exp(ikR)/R, \quad b = k(i - \cot\delta_0) = -1/f_0(k) \text{ and } d = -\exp(i\mathbf{k}\cdot\mathbf{R}/2), \qquad (16)$$

where $f_0(k)$ is the partial amplitude of electron elastic scattering by single atomic potential. Then (15) transforms into a system of two simple equations, $aD_1 + bD_2 = d$, $bD_1 + aD_2 = d^*$ the solution of which give the following expressions [35] for the coefficients at spherical wave (13)

---

[3] Of course, inside the atom $\psi$ is not divergent, but we consider low enough electron momentum $k$, so that the atomic sphere can be treated as zero radius object.

$$D_1(\mathbf{k}) = \frac{ad - bd^*}{a^2 - b^2} \quad \text{and} \quad D_2(\mathbf{k}) = \frac{ad^* - bd}{a^2 - b^2}, \tag{17}$$

The wave function $\psi_\mathbf{k}^+(\mathbf{r})$ (13) *beyond the atomic spheres* is the *exact general solution* of the Schrödinger equation that describes the multiple scattering of a particle by two-center target [35]. Indeed, if we substitute function (13) in the wave equation we obtain in its right side the sum of two delta-functions $\delta(\mathbf{r} \pm \mathbf{R}/2)$, because the second and third terms in (13) coincide up to constant with the Green functions of the wave equation for the free particle motion. Delta-functions are equal to zero in all space except the points $\mathbf{r} = \pm\mathbf{R}/2$. So, beyond the atomic spheres (where $\mathbf{r} \neq \pm\mathbf{R}/2$) the function (13) obeys the wave equation for free electron motion.

The amplitude of slow electron scattering by the target is obtained by considering the asymptotic behavior of the wave function (13). As result, we obtain the following *exact amplitude* of the electron multiple scattering by given two-center target [34, 35]

$$F(\mathbf{k}, \mathbf{k}', \mathbf{R}) = \frac{2}{a^2 - b^2} \{b \cos[(\mathbf{k} - \mathbf{k}') \cdot \frac{\mathbf{R}}{2}] - a \cos[(\mathbf{k} + \mathbf{k}') \cdot \frac{\mathbf{R}}{2}]\}. \tag{18}$$

The differential cross section of this process is given by following formula

$$\frac{d\sigma}{d\Omega_{k'}} = |F(\mathbf{k}, \mathbf{k}', \mathbf{R})|^2. \tag{19}$$

The total scattering cross section we obtain from the amplitude (19) with the help of the optical theorem [30]

$$\sigma(\mathbf{k}, \mathbf{R}) = \int \frac{d\sigma}{d\Omega_{k'}} d\Omega_{k'} = \frac{4\pi}{k} \operatorname{Im} F(\mathbf{k} = \mathbf{k}', \mathbf{R}) = \frac{4\pi}{k} \operatorname{Im}\left[\frac{b - a\cos(\mathbf{k} \cdot \mathbf{R})}{a^2 - b^2}\right]. \tag{20}$$

The presence of vectors $\mathbf{k}$ and $\mathbf{R}$ in the argument of the cross section (20) emphasize that we are dealing with fixed-in-space molecule. The total cross section (20) averaged over all the directions of the incident electron momentum $\mathbf{k}$ relative to the vector $\mathbf{R}$ has the form:

$$\bar{\sigma}(k) = \frac{1}{4\pi} \int \sigma(k, \mathbf{R}) d\Omega_k = \frac{8\pi}{k} \operatorname{Im}\left[\frac{b - a j_0(kR)}{a^2 - b^2}\right]. \tag{21}$$

Here $j_0(x) = \sin x / x$. Using the explicit expressions for the functions $a$ and $b$, we obtain the following formula for the averaged cross section [35]

$$\bar{\sigma}(k) = \frac{4\pi}{k^2} \left\{ \left[1 + \left(\frac{qR + \cos kR}{kR + \sin kR}\right)^2\right]^{-1} + \left[1 + \left(\frac{qR - \cos kR}{kR - \sin kR}\right)^2\right]^{-1} \right\}, \tag{22}$$

where $q = -k \cot \delta_0(k)$. The total cross section (22) contains $\sin kR / kR$ that characterizes diffraction phenomena. Its appearance is a consequence of the interference of two *s*-waves in the continuum wave function (13).

### 3.1 Expansion of the scattering amplitude in spherical functions

Let us represent the obtained before scattering amplitude (18) in spherical functions, like (2):

$$\frac{2}{a^2-b^2}\{b\cos[(\mathbf{k}-\mathbf{k}')\cdot\frac{\mathbf{R}}{2}]-a\cos[(\mathbf{k}+\mathbf{k}')\cdot\frac{\mathbf{R}}{2}]\}=\frac{2\pi}{ik}\sum_{l,m}(e^{2i\eta_l}-1)Y_{lm}^*(\mathbf{k})Y_{lm}(\mathbf{k}'). \quad (23)$$

Multiplying both parts of this equality by spherical functions and integrating over all directions of vectors **k** and **k'** we obtain the following equation for the *infinite number* of molecular phases $\eta_l(k)$:

$$e^{i\eta_l}\sin\eta_l = 8\pi k j_l^2(kR/2)|Y_{lm}(\mathbf{R})|^2\frac{b-(-1)^l a}{a^2-b^2}. \quad (24)$$

Here $j_l(x)$ is the spherical Bessel function. From the equality of the real and imaginary parts of the equation (24) we obtain for molecular phases the following expressions

$$\cot\eta_l = -\frac{qR+\cos kR}{kR+\sin kR}, \text{ for even } l, \quad \cot\eta_l = -\frac{qR-\cos kR}{kR-\sin kR}, \text{ for odd } l. \quad (25)$$

It is evident that the total cross section of scattering by two centers calculated with these phases (owing to their independence on *l*) diverges. Indeed, rewriting the sum of partial cross sections as two infinite sums, we obtain

$$\overline{\sigma}(k) = \frac{4\pi}{k^2}\sum_{l=0}^{\infty}(2l+1)\sin^2\eta_l = \frac{4\pi}{k^2}\left[\sum_{l=even}^{\infty}(2l+1)\sin^2\eta_l + \sum_{l=odd}^{\infty}(2l+1)\sin^2\eta_l\right] = \infty. \quad (26)$$

The reason for this meaningless result is the straightforward application of the usual S-matrix method for spherically symmetrical potential (formula (2)) to a target that is non-spherical. Thus, the amplitude of elastic scattering by a two-center target, *in principle, cannot be represented as a partial expansion in a series of spherical functions*. Therefore, the the methods for calculating the molecular phases described in [24]-[27] cannot be considered as justified and reliable.

### 3.2 Expansion of the amplitude in a series of functions $Z_\lambda(\mathbf{k})$

According to formulas (18) - (21), the amplitude (18) should be represented as a partial expansion in a series of functions $Z_\lambda(\mathbf{k})$. For given molecular system that is created by two short-range potentials, each being a source of the scattered *s*-waves only, the molecular phase shifts $\eta_\lambda(k)$ and the functions $Z_\lambda(\mathbf{k})$ can be calculated explicitly [35]. Let us rewrite the scattering amplitude (18) in another form

$$F(\mathbf{k},\mathbf{k}',\mathbf{R}) = -\frac{2}{a+b}\cos(\mathbf{k}\cdot\mathbf{R}/2)\cos(\mathbf{k}'\cdot\mathbf{R}/2)+\frac{2}{a-b}\sin(\mathbf{k}\cdot\mathbf{R}/2)\sin(\mathbf{k}'\cdot\mathbf{R}/2). \quad (27)$$

According to [29], the amplitude (27) should be considered as a sum of two partial amplitudes. The first of them is written as

$$\frac{4\pi}{2ik}(e^{2i\eta_0}-1)Z_0(\mathbf{k})Z_0^*(\mathbf{k'}) = -\frac{2}{a+b}\cos(\mathbf{k}\cdot\mathbf{R}/2)\cos(\mathbf{k'}\cdot\mathbf{R}/2). \tag{28}$$

The second one is defined by the following expression

$$\frac{4\pi}{2ik}(e^{2i\eta_1}-1)Z_1(\mathbf{k})Z_1^*(\mathbf{k'}) = \frac{2}{a-b}\sin(\mathbf{k}\cdot\mathbf{R}/2)\sin(\mathbf{k'}\cdot\mathbf{R}/2). \tag{29}$$

The reasons for assigning the indexis to the functions $Z_\lambda(\mathbf{k})$ the values $\lambda = 0, 1$ will become understandable as we proceed. After elementary transformations of the formulas (28) and (29), we obtain two molecular scattering phases ("proper phases" in [35])

$$\cot\eta_0 = \frac{\text{Re}[(a+b)^*]}{\text{Im}[(a+b)^*]} = -\frac{qR+\cos kR}{kR+\sin kR}, \quad \cot\eta_1 = \frac{\text{Re}[(a-b)^*]}{\text{Im}[(a-b)^*]} = -\frac{qR-\cos kR}{kR-\sin kR}. \tag{30}$$

Substituting the phase shifts (30) into formulas (28) and (29), we obtain the functions $Z_\lambda(\mathbf{k})$ in the explicit form. They are defined by the following expressions [35] (in [29] functions $Z_\lambda(\mathbf{k})$ are called the "characteristic scattering amplitudes")

$$Z_0(\mathbf{k}) = \frac{\cos(\mathbf{k}\cdot\mathbf{R}/2)}{\sqrt{2\pi S_+}}, \quad Z_1(\mathbf{k}) = \frac{\sin(\mathbf{k}\cdot\mathbf{R}/2)}{\sqrt{2\pi S_-}}. \tag{31}$$

Here $S_\pm = 1 \pm j_0(kR)$. It is easy to demonstrate that the functions (31) obey the conditions (7). Evidently, the functions (31) are defined by the geometrical target structure, i.e. by the direction of the molecular axis $\mathbf{R}$ in the arbitrary coordinate system, in which the electron momentum vectors before and after scattering are $\mathbf{k}$ and $\mathbf{k'}$, respectively. The transition to the limit $k \to 0$ in formulas (31) gives instead of the functions $Z_\lambda(\mathbf{k})$ the well-known spherical functions

$$Z_0(\mathbf{k})_{k\to 0} \to \frac{1}{\sqrt{4\pi}} \equiv Y_{00}(\mathbf{k}), \quad Z_1(\mathbf{k})_{k\to 0} \to \sqrt{\frac{3}{4\pi}}\cos\vartheta \equiv Y_{10}(\mathbf{k}). \tag{32}$$

Here $\vartheta$ is the angle between the vector $\mathbf{k}$ and axis $\mathbf{R}$. *Only in the limit $k \to 0$ it becomes correct to replace a non-spherical molecular field by a spherical one.*

The molecular phases $\eta_\lambda(k)$ in (30) can be classified by considering their behavior at $k \to 0$ [29]. In this limit the electron wavelength is much greater than the target size and the picture of scattering should approach that in the case of spherical symmetry. Considering the transition to this limit in the formulas (30), we obtain: $\eta_0(k \to 0) \sim k$ and $\eta_1(k \to 0) \sim k^3$. Thus, the molecular phases behave similar to the *s*- and *p*- phases in the spherically symmetrical potential, which explains the choice of their indexes.

### 4. Wigner time delay
Developed in Section 3.2 method of partial waves for non-spherical targets separates in a natural way the kinematics of the scattering process, which is defined by functions $Z_\lambda(\mathbf{k})$, from its dynamical part that determines the phase shifts $\eta_\lambda(k)$. The scattering amplitude (18) is determined by two partial waves with indexes λ = 0 and 1, which have the following asymptotic forms [35]

$$R_{k0}(r \to \infty) = e^{i\eta_0} \frac{1}{kr} \sin(kr + \eta_0); \qquad R_{k1}(r \to \infty) = e^{i(\eta_1 + \frac{\pi}{2})} \frac{1}{kr} \sin(kr - \frac{\pi}{2} + \eta_1). \qquad (33)$$

These two functions determine the particle fluxes through the surface of a sphere with a large radius surrounding the molecule. According to [13], these flows determine the following two delay times of the electron by the target. The first Wigner time is:

$$\tau_0(k) = 2\frac{d\eta_0}{dE} = \frac{2}{k}\dot{\eta}_0(k) = -\frac{2}{k} RF_0(k)[(qR + \cos kR)^2 + (kR + \sin kR)^2]^{-1} \qquad (34)$$

where function $F_0(k)$ has form

$$F_0(k) = (1 + \cos kR)(qR + \cos kR) - (\dot{q} - \sin kR)(kR + \sin kR). \qquad (35)$$

The second time delay is

$$\tau_1(k) = 2\frac{d\eta_1}{dE} = \frac{2}{k}\dot{\eta}_1(k) = -\frac{2}{k} RF_1(k)[(qR - \cos kR)^2 + (kR - \sin kR)^2]^{-1} \qquad (36)$$

where function $F_1(k)$ is of the form

$$F_1(k) = (1 - \cos kR)(qR - \cos kR) - (\dot{q} + \sin kR)(kR - \sin kR). \qquad (37)$$

In these formulas the functions $q(k)$ and $\dot{q}(k)$ are

$$q(k) = -k \cot \delta_0(k); \qquad \dot{q} = -\frac{\sin \delta_0 \cos \delta_0 - k\dot{\delta}_0}{\sin^2 \delta_0}. \qquad (38)$$

Operator $\partial/\partial k$ in formulas (34)-(38) is denoted by a dot.

Fig. 1 presents the times delay $\tau_0(k)$ and $\tau_1(k)$ for molecule $C_2$ in the model of two non-overlapping atomic spheres for four fixed values of inter-atomic distances $R_C$. In the real molecule $C_2$ this distance is $R_C$=2.479 atomic units (au) [35]. In these calculations with formulas (34)-(38) we describe the atomic s-phase shift $\delta_0(k)$ by the following analytical expression $\delta_0(k) \approx 2\pi - 1.912k$ [35]. Time delay $\tau_0(k)$ goes to minus infinity at electron wave vector $k$ goes to zero

$$\tau_0(k \to 0) \cong \frac{2}{k}\frac{d}{dk}(2\pi - 1.912k) = -3.824/k. \qquad (39)$$

We can estimate the ability of an electron to penetrate the potential field of a molecule by observing the behavior of the second time delay $\tau_1(k)$. According to Fig. 1, with increase of interatomic distance $R_C$, the minimum value of the function $\tau_1(k)$ increases i.e. an increase in the interatomic distance $R_C$ of the molecule is accompanied by a more intense expulsion of the electron from the inner region of the molecule. A similar picture is given in the article [16] in Fig. 4, which shows the dependence of the photoelectron delay time by $H_2$ molecule as a function of the interatomic distance. The time delay becomes negative at $R > 2.6$ au and becomes deeper with the growth of $R$. Both curves in Fig. 1 in this case rapidly oscillate, which is associated with the presence in the scattering phases of terms $\sin kR / kR$ that are typical for interference of two $s$-waves in the continuum wave function (13).

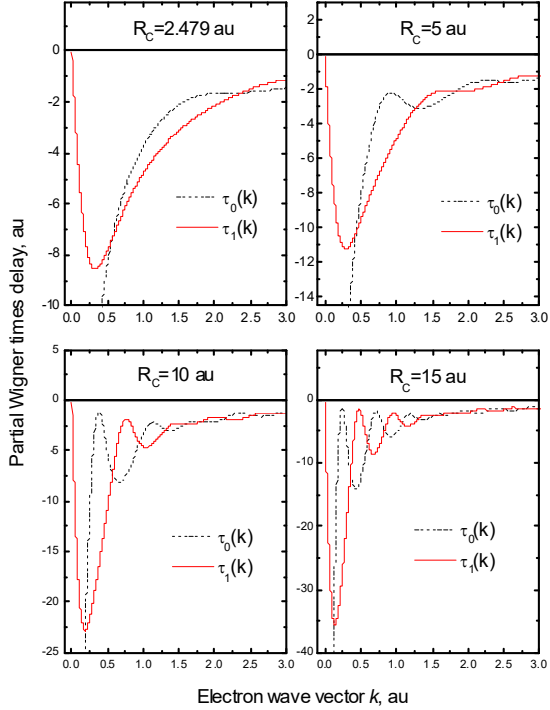

Fig. 1. Wigner times delay for $C_2$ molecule versus electron wave vector $k$ calculated within the model of non-overlapping atomic spheres. $R_C$ is the inter-atomic distance.

## 5. Conclusion

In studies of electron-molecule scattering, the continuum electron usually is treated as moving in a spherically averaged molecular field. The wave functions describing the scattering of an electron by a polyatomic molecule outside the so-called molecular sphere are considered as a linear combination of regular and irregular solutions of the wave equation for free space (in the case of neutral target) or the Coulomb functions for molecular ions. The phase shifts of molecular wave function are defined from the matching condition for the solutions of the wave equation inside and outside this sphere. The purpose of this article was to analyze whether this method to find the molecular phases is correct. It was shown that a straightforward application of the classical S-matrix method to non-spherical potentials leads to irreparable internal contradictions and therefore this approach to calculation of the scattering phases cannot be considered as satisfactory. We came to this conclusion based on the consideration of the simplest model system formed by two non-overlapping atomic potentials. However, we believe that, as in the case of two-atomic system, the amplitude of elastic scattering of a particle on any multicenter target of this type can be represented as an expansion in a series of orthonormal functions $Z_\lambda(\mathbf{k})$ specific to a given target ("characteristic scattering amplitudes" [29]). The coefficients of this expansion determine a unique (again for a given target) set of molecular scattering phases $\eta_\lambda(k)$ ("proper phase shifts" [29]). These phases also determine the Wigner times delay of an electron by a given molecule. We do believe that the developed approach will open new horizon in studies of electron-molecule collisions, including their temporal picture.

**Acknowledgments**

The Uzbek Foundation Award OT-Ф2-46 supported the work of ASB.

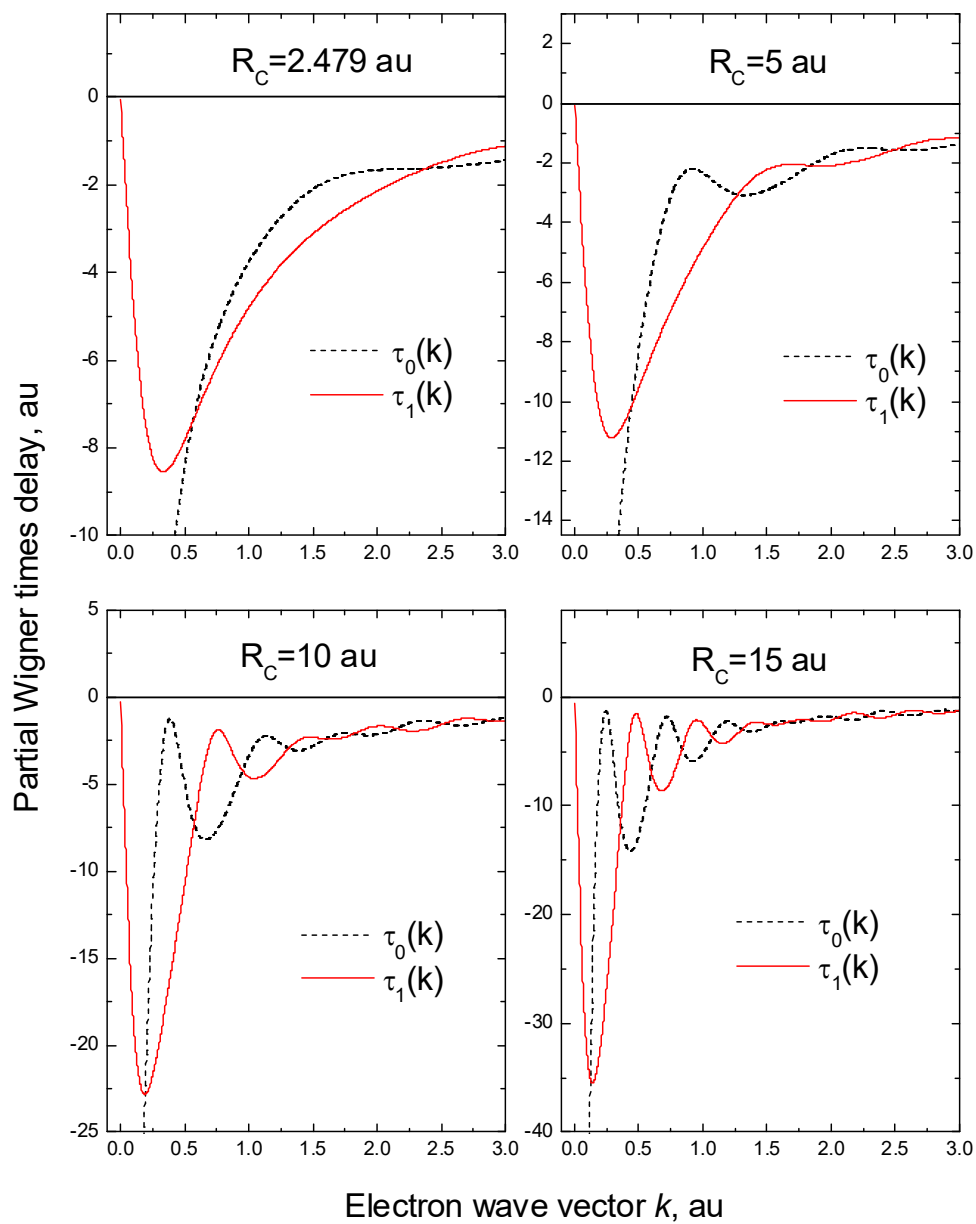

Fig. 1. Wigner times delay for $C_2$ molecule versus electron wave vector $k$ calculated within the model of non-overlapping atomic spheres. $R_C$ is inter-atomic distance.